\documentclass[twocolumn,showpacs,preprintnumbers,amsmath,amssymb,floatfix,nofootinbib]{revtex4}

\usepackage{graphicx}
\usepackage{dcolumn}
\usepackage{bm}

\def \pom {{I\!\!P}}

\begin{document}

\title{Investigating the central diffractive $f_0(980)$ and $f_2(1270)$ meson production at the LHC}

\author{M. V. T. Machado}
\affiliation{High Energy Physics Phenomenology Group, GFPAE  IF-UFRGS \\
Caixa Postal 15051, CEP 91501-970, Porto Alegre, RS, Brazil}

\begin{abstract}
The central diffractive production of mesons $f_0(980)$ and $f_2(1270)$ at the energy of CERN-LHC experiment on proton-proton collisions is investigated. The processes initiated by quasi-real photon-photon collisions and by central diffraction processes are considered. The role played by the photon-Odderon production channel is also studied. The cross sections for these distinct production channels are compared and analyzed.
\end{abstract}


\maketitle

\section{Introduction}

In the last two years, ALICE collaboration has recorded zero bias and minimum bias data in proton-proton collisions at a center-of-mass energy of $\sqrt{s}=7$ TeV. Among the relevant events, those containing double gap topology have been studied and they are associated to central diffractive processes \cite{ALICE}. In particular, central meson production was observed. In the double gap distribution, the $K_s^0$ and $\rho^0$ are highly suppressed while the $f_0(980)$ and $f_2(1270)$ with quantum numbers $J^{PC}=(0,2)^{++}$ are much enhanced. Such a measurement of those states is evidence that the double gap condition used by ALICE selects events dominated by double Pomeron exchange. The central exclusive production processes in high energy collisions have traditionally considered a promising way to study particles in a especially clean environment in which to measure the nature and quantum numbers, e.g. spin and parity, of the centrally produced state.  In addition they give information about the structure of the Pomeron and of the mechanism of Pomeron-Pomeron interaction. 

From the experimental point of view, there are much activity in the study of these exclusive meson production processes at the Tevatron \cite{TevCHI}, where exclusive $\chi_c$ production has been investigated. Those measurements are in broad agreement with theoretical expectations \cite{KhozeCHI,AntoniCHI}. Due to the reasonably large mass of quarkonia perturbative QCD methods can be considered for theoretical calculations. For the light meson production the situation is not so clear. For instance, some attempts have been done in describing $f_0(1500)$ in a QCD perturbative framework \cite{Antonif0}. An additional complication in the subject is the possibility of production of exotic mesons. The gluon self-coupling in QCD opens the possibility of existing bound states of pure gauge fields known as glueballs \cite{Vento,Crede}. Eventually, this glueball state will mix strongly with nearby $q\bar{q}$ states \cite{closekirk,amsler}.  Many mesons  have stood up as good candidates for the lightest glueball in the spectrum and in particular the scalar sector ($J^{PC} = 0^{++}$) seems promising. The identification of glueballs \cite{Ochs1} is not easy and some methods have been introduced during last years. Recently, it is expected that studies on gluon jets at the LHC could find glueballs in a clear way \cite{Ochs2}.
 
Last year, the first heavy ion run at the LHC took place and data were taken with two dedicated triggers for investigating meson production in the ALICE central barrel.  Ongoing studies \cite{ALICE} report that the production of $\rho^0$-meson with $J^{PC}=1^{--}$ has been measured and indicates that the double gap events in PbPb collisions are not dominated by Pomeron-Pomeron events as is the case in proton-proton collisions. In fact, the dominant photon-Pomeron production channel results in the diffractive photoproduction of vector mesons like $\rho$ \cite{Machadofoto} and photon-photon production channel results in electromagnetic production of pseudo-scalars like $\pi$'s and $\eta$'s. The cross sections for these processes are smaller than the correspondent inclusive production channels, which it is compensated by a more favorable signal/background relation. In Ref. \cite{Mario}, we investigated the exclusive production of exotic meson in two-photon and Pomeron-Pomeron interactions in coherent PbPb collisions at the LHC. The photon flux scales as the square charge of the beam, $Z^2$, and then the corresponding cross section is highly enhanced by a factor $\propto Z^4\approx 10^7$ for Pb compared to proton case. A production channel producing similar final state configuration is the central diffraction process which is is modeled in general by two-Pomeron interaction. Compared to the two-photon process, the  Pomeron-Pomeron calculations are subject to large uncertainties at collider energies. In Ref. \cite{Mario} the cross sections for these two channels were compared for the exotic meson production.

Here, we will focus on the central diffractive production of mesons $f_0(980)$ and $f_2(1270)$ in proton-proton collisions at LHC energies. The present investigation is relevant for the ATLAS, CMS and ALICE experiments. We will consider the Pomeron-Pomeron processes and investigate their theoretical uncertainties. In particular, we consider nonperturbative Pomeron model which is justified by the low mass of considered resonances. For completeness, we compute the cross sections rising from the two-photon channel. Finally, we investigate the size of the photon-Odderon contribution. Namely, we study the implication of high-energy photoproduction of $C=+1$ mesons as $f_2(1270)$ with nucleon excitation through Odderon exchange \cite{Odderon}. This paper is organized as follows: in next  section we present the main expressions for cross section calculation of two-Pomeron and two-photon processes and in last section we shown the numerical results and discussions. In the last section we also estimate the contribution of the Odderon exchange.

\section{Cross section calculation}

Let us start with the exclusive meson production from central diffractive reactions. That is the Pomeron-Pomeron production channel. In particular, we focus on the central diffraction (double Pomeron exchange, DPE) in proton-proton interactions. Due to the light mass of mesons $f_0(980)$ and $f_2(1270)$ a non-perturbative approach for the Pomeron exchange should be reasonable. We first focus on the isoscalar meson $f_0(980)[0^{++}]$. We will start with the phenomenological model for the soft Pomeron \cite{Land-Nacht,Bial-Land}.  In the exclusive DPE event the central object $X$ is produced alone, separated from the outgoing hadrons by rapidity gaps,
$pp\rightarrow p+\text{gap}+X+\text{gap}+p$. In approach we are going to use, Pomeron exchange corresponds to the exchange of a
pair of non-perturbative gluons which takes place between a pair of colliding quarks. The scattering matrix is given by,
\begin{eqnarray}
\mathcal{M} & = & \mathcal{M}_{0}\left(  \frac{s}{s_{1}}\right)  ^{\alpha(t_{2})-1}\left(  \frac{s}{s_{2}}\right)
^{\alpha(t_{1})-1}\,F(t_{1})\,F(t_{2})\nonumber \\
&\times & \exp\left(  \beta\left(  t_{1}+t_{2}\right)  \right)\,  S_{\text{gap}}\left(\sqrt{s} \right).
\label{M_all}
\end{eqnarray}
Here $\mathcal{M}_{0}$ is the amplitude in the forward scattering limit ($t_1=t_2=0$). The standard Pomeron Regge
trajectory is given by $\alpha\left(  t\right)=1+\epsilon+\alpha^{\prime}t$ with
$\epsilon\approx 0.08,$ $\alpha^{\prime}=0.25$ GeV$^{-2}$. The momenta of incoming (outgoing) protons are labeled by $p_1$
and $p_2$ ($k_1$ and $k_2$), whereas the glueball momentum is denoted by $P$. Thus, we can define the following quantities
appearing in Eq. (\ref{M_all}): $s=(p_{1}+p_{2})^{2}$, $s_{1}=(k_{1}+P)^{2},$ $s_{2}=(k_{2}+P)^{2},$
$t_{1}=(p_{1}-k_{1})^{2}$ , $t_{2}=(p_{2}-k_{2})^{2}$. The nucleon form-factor is given by $F_p\left(  t\right)  $ = $\exp(b
t)$ with $b=$ $2$ GeV$^{-2}$. The phenomenological factor $\exp\left(  \beta\left(  t_{1}+t_{2}\right)  \right)
$ with $\beta$ $=$ $1$ GeV$^{-2}$ takes into account the effect of the momentum transfer dependence of the non-perturbative
gluon propagator. The factor $S_{\text{gap}}$ takes the gap survival effect into account $i.e.$ the probability
($S_{\text{gap}}^{2}$) of the gaps not to be populated by secondaries produced in the soft rescattering.  For our purpose
here, we will consider $S_{\mathrm{gap}}^2=0.026$ at $\sqrt{s}=14$ TeV in nucleon-nucleon collisions \cite{KKMR}. Such a value for the survival gap factor is typical for soft processes and should reasonable for a cross section estimation involving light mesons as $f_0$ (its mass scale is typical from the nonperturbative regime). However, it should be noticed that the particular values of $S_{\mathrm{gap}}^2$ is dependent on the mass and spin of centrally produced system. An updated discussion on those dependencies for the heavy $chi_{c,b}$ mesons can be found in Ref. \cite{KMRCHIS2}. When computing cross sections at 7 TeV we will use the interpolation value $S_{\mathrm{gap}}^2=0.032$.

Following the calculation presented in Ref. \cite{Bial-Land} we find
$\mathcal{M}_{0}$ for colliding hadrons,
\begin{eqnarray}
\mathcal{M}_{0}=32 \,\alpha_0^2\,D_{0}^{3}\,\int d^{2}\vec{\kappa}\,p_{1}^{\lambda}V_{\lambda\nu}^{J}p_{2}^{\nu}\,
\exp(-3\,\vec{\kappa }^{2}/\tau^{2}),
\label{M_o}
\end{eqnarray}
where $\kappa $ is the transverse momentum carried by each of the three gluons. $V_{\lambda\nu}^{J}$ is the
$gg\rightarrow R^{J}$ vertex depending on the polarization $J$ of the $R^{J}$ meson state. The fixed parameters of model are $\tau=1$ GeV and $D_{0}G^{2}\tau=30$ GeV$^{-1}$ \cite{Bial-Land} where $G$ is the scale
of the process independent non-perturbative quark gluon coupling.  We consider the parameter
$\alpha_0=G^2/4\pi$ as free and it has been constrained by the experimental result for $\chi_{c}\,(0^{++})$ production at Tevatron \cite{TevCHI}, $d\sigma\,(\chi_{c0})/dy|_{y=0}=76\pm 14 $ nb.  Namely, we found the constraint $S_{\mathrm{gap}}^2\,(\sqrt{s}=2\,\mathrm{TeV})/\alpha_0^2 = 0.45$, where $S_{\mathrm{gap}}^2$ is the gap survival probability factor (absorption factor). Considering the KMR \cite{KKMR} value $S_{\mathrm{gap}}^2=0.045$ for central diffractive processes at Tevatron energy, one obtains $\alpha_0=0.316$. We notice that the CDF collaboration \cite{TevCHI} assumes the absolute dominance of the spin-$0$ contribution in the charmonium production in the radiative $J/\Psi+\gamma$ decay channel (the events had a limited mass resolution and were collected in a restricted area of final-state kinematics)  and then the result was published as:
\begin{eqnarray}
\left.\frac{d\sigma[\chi_c(0^+)]}{dy}\right|_{y=0}& \simeq & \frac{1}{\mathrm{BR}[\chi_c(0^+)]}\left.\frac{d\sigma[pp\rightarrow pp(J/\Psi+\gamma)]}{dy}\right|_{y=0}\nonumber \\
&=& (76\pm 14)\,nb,
\end{eqnarray}
where $\mathrm{BR}[\chi_c(0^+)]=\mathrm{BR}(\chi_c(0^+)\rightarrow J/\Psi+\gamma)$ is the corresponding branching ratio. This fact is not true for general kinematics, as indicated by the Durham group investigations \cite{KMRCHIS2}.

For the
isoscalar meson $f_0(980)$ considered here, $J=0$, one obtains the following result
\cite{Bial-Land,KMRS}:
\begin{equation}
p_{1}^{\lambda}V_{\lambda\nu}^{0}p_{2}^{\nu}=\frac{s\,\vec{\kappa}^{2}
}{2M_{G^{0}}^{2}}A, \label{p_1Vp_2}%
\end{equation}
where $A$ is expressed by the mass $M_{G}$ and the width $\Gamma (gg\rightarrow R)$ of the meson resonance $R$ through the relation:
\begin{equation}
A^{2}= 8\pi M_R\,\Gamma (gg\rightarrow R). \label{A^2}
\end{equation}
For obtaining the two-gluon decays widths the following relation is used,
$\Gamma \,(R\rightarrow gg)=\mathrm{Br}\,(R\rightarrow gg)\,\Gamma_{tot}(R)$. For simplicity, we will take $\mathrm{Br}\,(R\rightarrow gg)=1$, which will introduce a sizable theoretical uncertainty. The two-gluon width depends on the branching fraction of the resonance $R$ to gluons. It is timely to mention that for scalar mesons which are glueballs candidates, is a theoretical expectation \cite{Farrar} that $\mathrm{Br}\,(R(q\bar{q})\rightarrow gg)={\cal O}(\alpha_s^2)\simeq 0.1-0.2$ whereas $\mathrm{Br}\,(R(G)\rightarrow gg)\simeq 1$.  The values for $\Gamma_{gg}$ used in our calculations are summarized in Table I. The numerical results for the LHC energies in proton-proton collisions are presented also in the Table I. The rapidity distribution $d\sigma (y=0)/dy$ and the total integrated cross section are computed. These values can be compared to previous calculations including production in the heavy-ion mode summarized in Ref. \cite{hep_lowx}.

A limitation of the approach above is that it does not allow to deal with $J=1,2$ states.  This is the case for the meson $f_0(1270)$, which is a state $J^{PC}=2^{++}$. It has been shown \cite{Yuan} that the DPE contribution to $J=1$ and $J=2$ meson production in the forward scattering limit is vanishing, either perturbative or non-perturbative Pomeron models. This limitation can be circumvented if we consider Donnachie-Landshoff Pomeron \cite{DOLA}, where it is considered like a isoscalar ($C=+1$) photon when coupling to a quark or anti-quark. In this approach, the cross section is written as:
\begin{eqnarray}
&& \sigma \,(pp\rightarrow p+R+p)  =   \frac{1}{2(4\pi )^3s^2W_{R_J}^2}\int dP_{R} \,dt_1 dt_2 \nonumber \\
 & & \times  \sum_{j=1}^{2}\omega_j\ell_1^{\mu \alpha} \ell_2^{\nu \beta}\,A_{\mu \nu}^JA_{\alpha \beta}^{J*}\left[D_{\pom}(t_1)D_{\pom}(t_2) \right]^2.
\end{eqnarray}

Here, the effective Pomeron propagator is giving by,
\begin{eqnarray}
D_{\pom}(t)=3\beta_0^2\left(\frac{\omega }{E}\right)^{1-\alpha_{\pom}(t)}\,F_p(t),
\end{eqnarray}
where $\beta_0=1.8$ GeV$^{-1}$ and the form factor $F_p(t)$ of the nucleon is taken into account in the form of
\begin{eqnarray}
F_p(t) = \frac{4m_p^2-2.8t}{4m_p^2-t}\left(1-\frac{t}{0.7\mathrm{GeV}^2}\right)^{-2}.
\end{eqnarray}
The coupling to the nucleon is described by the tensor $\ell^{\mu \alpha}$ arising from its fermionic current. For the Pomeron-energies it is used $\omega_{1,2}=(W_{R_J}\pm P_{R})/2$, where $W_{R_J}$ corresponds to the total energy of the meson $R_J$ in the center-of-mass system given by $W_{R_J} = P_{R}^2+M_R^2$. Concerning the Pomeron-Pomeron-R vertex, the $R$ particle is treated as a non-relativistic bound state of a $q\bar{q}$ system. Since the Pomeron couples approximately like a $C=+1$ photon, the Pomeron-quark vertex is given by a $\gamma$-matrix. For the amplitude, we show the explicit formulae:
\begin{eqnarray}
A_{\mu \nu}^{J=0} & = & A_0\left\{ \left[g_{\mu \nu}(q_1\cdot q_2)-q_{2\mu}q_{1\nu} \right]\left[M_R^2+  (q_1\cdot q_2)\right]\right. \nonumber \\
& - & \left. g_{\mu\nu}q_1^2q_2^2 \right\},\nonumber \\
A_{\mu \nu}^{J=1} & =&  A_1\left(q_1^2\epsilon_{\alpha \mu \nu \beta} \epsilon^{\alpha}q_2^{\beta}- q_2^2\epsilon_{\alpha \mu \nu \beta} \epsilon^{\alpha}q_1^{\beta}\right),\nonumber \\
A_{\mu \nu}^{J=2} & = & A_2\left[(q_1\cdot q_2)g_{\mu \rho}g_{\nu \rho}+g_{\mu\nu}q_{1\rho}q_{2\sigma} \right. \nonumber \\
& - & \left. q_{2\mu}q_{1\rho}g_{\sigma\nu}-q_{1\nu}q_{2\rho}g_{\sigma \mu}  \right]\epsilon^{\rho \sigma}, \nonumber
\end{eqnarray}
where $A_0 = \frac{2}{\sqrt{6}}\frac{a}{M_R}$, $A_1 = ia$ and $A_2=\sqrt{2} a M_R$. The formulae above have been firstly obtained in Ref. \cite{KKS} for photon-photon fusion into a quarkonium state. In addition, $\epsilon_{\mu}$ and $\epsilon_{\mu \nu}$ are the polarization vector and tensor of the $J=1$ and $J=2$ states, respectively. The factor $a$ is given by:
\begin{eqnarray}
a=\frac{4}{(q_1\cdot q_2)}\sqrt{\frac{6}{4\pi\,M_R}}\,\phi^{\prime}(0)\,
\end{eqnarray}
where $\phi^{\prime}(0)$ denotes the derivative of the wavefunction at the origin in coordinate space, which can be determined from meson two-photon width  $\Gamma(R_{J=2}\rightarrow \gamma\gamma)$.

As a cross check, we have applied the approach above for computing the charmonium cross section. Namely, the derivative of the $P$-wave spatial wave function at the origin for the $\chi_c(J=2)$ is given by \cite{Lansberg},
\begin{eqnarray}
\Gamma_{\gamma \gamma}(\chi_{c2}) & = & \left(\frac{4}{15}\right)\frac{4\pi Q_c^4\alpha^2f_{\chi_{c0}}^2}{M_{\chi_{c2}}}\left[1+B_2\left(\frac{\alpha_s}{\pi}\right) \right],\\
f_{\chi_{c0}} & = & 12 \,\sqrt{\frac{3}{(8\pi m_Q)}}\left(\frac{\phi^{\prime}(0)}{M} \right),
\end{eqnarray}
where $Q_c=2/3$ and $B_2=-16/3$ is the next-to-leading-order (NLO) QCD radiative correction. Here, we have neglect the binding energy and set $M_R=2m_Q$ and also take $\alpha_s=0.316$. The calculation for the Tevatron energy, $\sqrt{s} = 1.96$ TeV gives, $\sigma_{tot}(\chi_{c0}) = 96$ nb, $\sigma_{tot}(\chi_{c1}) = 3.0$ nb and $\sigma_{tot}(\chi_{c2}) = 28$ nb, respectively. This is order of magnitude consistent with the Tevatron data \cite{TevCHI}. It can be compared to the previous results in literature: the Krakow Group computed recently \cite{Krakow2010}  the central exclusive production of $chi_c$, which gives $\sigma[\chi_c(0^+)]\simeq 97\pm 22$ nb, $\sigma[\chi_c(1^+)]\simeq 3.2\pm 0.1$ nb and $\sigma[\chi_c(2^+)]\simeq 4.3 \pm 1.7$ nb (we have included the theoretical uncertainty). The agreement is good with exception of $2^+$ state. Concerning the recent results from the Durham Group \cite{Durham2010}, they obtain $\sigma[\chi_c(0^+)]\simeq 130$ nb, $\sigma[\chi_c(1^+)]\simeq 79$ nb and $\sigma[\chi_c(2^+)]\simeq 28$ nb (we have assumed $\Delta y=2$ as only the distribution $d\sigma/dy (y=0)$ is predicted in \cite{Durham2010}).

We focus now on the estimate for the $f_2(1270)$ meson. The non-relativistic quark model  predicts that its two-photon partial width is given by \cite{Barnes},
\begin{eqnarray}
\Gamma_{\gamma \gamma}(f_2(1270)) = 3\left(\frac{5}{9\sqrt{2}}\right)^2\frac{12}{5}\frac{2^4\alpha^2}{M^4}|\phi^{\prime}(0)|^2.
\end{eqnarray}
Our numerical results for the LHC energy are presented in Table II. The prediction for Tevatron at $\sqrt{s}=1.96$ TeV is $\sigma_{tot}(f_2)=1058 $ nb. We have checked that the result agrees in order of magnitude with the CERN WA102 data \cite{WA102} at $\sqrt{s}=29.1$ GeV (using $S_{\text{gap}}^2=1$). We found $\sigma_{th}=5130$ nb versus the experimental value $\sigma_{exp} = 3275\pm 422$ nb.
We have computed the $f_0(980)$ cross section at $\sqrt{s}=7$ TeV in the approach above, giving $\sigma_{tot}(f_0(980))\simeq 10$ $\mu$b, which is at least one order of magnitude below the soft Pomeron model (see Table I).

\begin{table}[t]
\centering
\renewcommand{\arraystretch}{1.5}
\begin{tabular}{c c c c}
\hline
 $f_0(980)$ & $\Gamma_{tot}$ [MeV] & $\sqrt{s}=7$ TeV & $\sqrt{s}=14$ TeV \\
\hline
  $\frac{d\sigma}{dy}(y=0)$  & ($70\pm 38$)   & 26.9 $\mu$b  & 27.1 $\mu$b \\
  $\sigma_{tot}$  & --- &  369 $\mu$b &  407 $\mu$b \\
  \hline
\end{tabular}
\caption{Rapidity distribution at $y=0$ and integrated cross sections for isoscalar meson $f_0(980)$ in the soft Pomeron model at the LHC energies.}
\label{tab1}
\end{table}

Now, we analyze the the two-photon production channel. The high energy photon-induced interaction with exclusive meson final state occurs when two quasi-real photons emitted by each proton interact with each other to produce the meson resonance  $\gamma \gamma \rightarrow R$. Deflected protons and their energy loss can be detected by forward detectors, but meson will go to the central detector. Photons emitted with small angles by the protons show a spectrum of virtuality $Q^2$ and energy $E_{\gamma}$.  This is described by the equivalent photon approximation \cite{EPA},
\begin{eqnarray}
\frac{dN}{dE_{\gamma}dQ^2} & =  &  \frac{\alpha}{\pi}\left[\left(1-\frac{E_{\gamma}}{E} \right)\left( 1-\frac{Q_{min}}{Q^2} \right)F_E \right. \nonumber \\
& + & \left. \frac{E_{\gamma}^2}{2E^2}F_M \right]\frac{1}{E_{\gamma}\,Q^2}.
\label{lumi1}
\end{eqnarray}

Here, $E$ is the energy of the proton beam which is related to the photon energy by $E_{\gamma} = xE$ and $m_p$ is the mass of the proton. For a collider, the center of mass energy is $\sqrt{s}=2E$. The functions $F_E$ and $F_M$ are the electric and magnetic form factors, The cross section $\sigma_{\gamma \gamma \rightarrow R}$ for the subprocess $\gamma \gamma \rightarrow R$ should be integrated over the photon spectrum 
\begin{eqnarray}
\sigma_{pp(\gamma\gamma)\rightarrow pRp}(\sqrt{s}) = \int \frac{dL^{\gamma \gamma}}{dW}(W,s)\,\sigma_{\gamma \gamma \rightarrow R}(W)\,dW,
\end{eqnarray}
where the effective photon luminosity is given by \cite{EPA},
\begin{eqnarray}
\frac{dL^{\gamma \gamma}}{dW} & = & \int_{Q_{1,min}^2}^{Q_{max}^2} dQ_1^2 \int_{Q_{2,min}^2}^{Q_{max}^2}dQ_2^2 \int_{y_{min}}^{y_{max}}dy \nonumber \\
&\times & \frac{W}{2y}\,f_1\left(\frac{W^2}{4y},Q_1^2\right)f_2\left(y,Q_2^2\right),
\label{lumi}
\end{eqnarray}
with
\begin{eqnarray}
Q_{min}^2 & = & \frac{m_p^2E_{\gamma}^2}{E(E-E_{\gamma})},\\
y_{min} & = & \mathrm{max}\left(\frac{W^2}{4x_{max}E},x_{min}E \right),\\
y_{max} & = & x_{max}E.
\end{eqnarray}

Here,$W$ is the invariant mass of the two-photon system $W_{\gamma \gamma}=2E\sqrt{x_1x_2}$ and the maximum virtuality is $Q_{max}^2=2$ GeV$^2$.
The function $f$ is defined by $f = dN/dE_{\gamma}dQ^2$.
\begin{table}[t]
\centering
\renewcommand{\arraystretch}{1.5}
\begin{tabular}{c c c c}
\hline
 $f_2(1270)$ & $\Gamma_{\gamma \gamma}/\Gamma_{tot}$  & $\sqrt{s}=7$ TeV & $\sqrt{s}=14$ TeV \\
\hline
  $\sigma_{tot}$  & $(1.64\pm 0.19)\times 10^{-5}$ & 1083 nb &  1107 nb \\
  \hline
\end{tabular}
\caption{Integrated cross sections for the meson $f_2(1270)$ in the Donnachie-Landshoff Pomeron  model at the LHC energies.}
\label{tab2}
\end{table}

The meson resonance production in two-photon fusion can be calculated using the narrow resonance approximation \cite{BKT}:
\begin{eqnarray}
\sigma\,(\gamma\gamma\rightarrow R) = (2J+1)\,\frac{8\pi^2}{M_R}\,\Gamma(R\rightarrow \gamma \gamma)\,\delta
\left(W_{\gamma\gamma}^2 -  M_R^2  \right),
\label{twophotres}
\end{eqnarray}
where $\Gamma(R\rightarrow \gamma \gamma)$ is the partial two-photon decay width, $M_R$ is the meson mass
and $J$ is the spin of the state $R$. Here, we compute the production rates for the mesons $f_{0}(980)$,
 and $f_2(1270)$, respectively.  The values for the corresponding widths were taken from the PDG average value \cite{PDG} and corresponding cross sections estimates are shown in Table III. By simple inspection we see that the two-photon channel is several orders of magnitude smaller than the Pomeron-Pomeron channel.

\begin{table}[t]
\centering
\renewcommand{\arraystretch}{1.5}
\begin{tabular}{l c c c}
\hline
 Meson & $\Gamma_{\gamma \gamma}$ [keV] & $\sqrt{s}=7$ TeV & $\sqrt{s}=14$ TeV \\
\hline
    $f_0(980)$ & ($ 0.29\pm 0.09$) & 0.12 nb & 0.15 nb \\
    $f_2(1270)$ & ($2.6\pm 0.24)$ & 2.57 nb & 3.37 nb \\
  \hline
\end{tabular}
\caption{Cross sections for the two-photon production channel at the LHC energies.}
\label{tab3}
\end{table}

In what follows we discuss the general features and uncertainties in the two referred production channels and in addition we discuss an estimate of $f_2(2170)$ produced in a photon-Odderon channel.

\section{Results and discussions}

In Table I the cross sections for scalar meson production in the soft Pomeron model  at the LHC energies are shown. The cross sections are reasonably large, despite the nonperturbative Pomeron  energy behavior of the considered model. The deviation is quite sizable when considering the Donnachie-Landshoff Pomeron, which gives  $\sigma_{f_0}(\sqrt{s}=7 \,\mathrm{TeV}) \approx 10$ $\mu$b. The deviation by a factor ten remains even accounting for a small branching ratio of scalar meson in two-gluons (we have used the simplification $\Gamma_{gg}=\Gamma_{tot}$). We quote Ref. \cite{Mario}, where implications from the meson content to the overall normalization of cross sections are discussed. The predictions for meson $f_2(1270)$ considering the Donnachie-Landshoff Pomeron at the LHC energy are presented in Table II, whereas we found for Tevatron the estimate $\sigma_{f_2}=1058 $ nb.

In Table III the results for two-photon production for both meson resonances $f_0$ and $f_2$ are presented. They are typically of order of a few nanobarns. As expected, they are are several orders of magnitude smaller than the Pomeron-Pomeron channel.  The present result is difficult to be compared directly to previous studies on Refs. \cite{Natale,Schramm}, where heavy-ions collisions have been considered. Concerning the overall normalization, for the scalar meson $f_0(980)$ the cross section could be even suppressed in case of considering it as a glueball candidate \cite{Mario}.

As a final analysis, we would like to address the photon-Odderon production channel in the specific case of the meson $f_2(1270)$. The phenomenological Odderon, a $C = P = -1$ partner of the $C = P = +1$ Pomeron, could exist \cite{Nicolescu}. Indeed within perturbative QCD, the Odderon is rather well defined with an
intercept $\alpha_{\mathrm{odd}}(0) \approx 1$ (for a review on QCD Odderon, see \cite{Ewerz}).  Applications in
the nonperturbative regime have assumed a ``maximal''
Odderon with an intercept $\alpha_{\mathrm{odd}}(0) \approx 1$. The exchange of the
phenomenological Odderon should produce a difference between $p p$ and
$\bar{p}p$ scattering at high energy and small momentum transfer, a
particularly sensitive test being provided by the forward real part of
the  $p p$ and $\bar{p}p$ scattering amplitudes. However, measurements
are consistent with the absence of odderon exchange \cite{Nicolescu}.
As an alternative, it was suggested
\cite{Nachtmann} that high-energy photoproduction of $C=+$ mesons, e.g.
$\pi^0$, $f_2^0(1270)$ and $a_2^0(1320)$, with nucleon excitation
would provide a clean signature for odderon exchange. In particular, it was theoretically predicted the following cross section at $\sqrt{s} = 20$ GeV for the $f_2(1270)$ meson \cite{Odderon}:
\begin{eqnarray}
\sigma(\gamma p \rightarrow f_2^0(1270) X) \approx 21\,{\mathrm{nb}}.
\label{predict}
\end{eqnarray}
whereas the the experimental results at $\sqrt{s} = 200$ GeV for $f_2(1270)$ \cite{H1f2m} is $\sigma(\gamma p \rightarrow f_2^0(1270) X) < 16$ nb at the 95$\%$ confidence level.

The model referred above is based on an approach to high-energy diffractive scattering
using functional integral techniques  and an extension of the model of the stochastic vacuum (for details, see Ref. \cite{Odderon}).  It is easily extended to Odderon exchange and gives an Odderon intercept $\alpha_{\mathrm{odd}}(0) = 1$. The scattering amplitude
$T(s,t)$ is obtained through a profile function $J(\vec b,s)$ :
\begin{eqnarray}
T(s,t) = 2is \int \, d^2b\;\exp (i\vec{q}\cdot\vec{b})\,J(\vec b, s).
\end{eqnarray}
The function $J(\vec b,s)$ is given in turn by the overlap of a dipole-dipole
scattering amplitude $\tilde{J}(\vec b, \vec r_1, \vec r_2, z_1, z_2)$
with appropriate wave functions for the initial and final states:
\begin{eqnarray}
J(\vec b, s) &=&-\int \frac{d^2r_1}{4\pi}dz_1\int \frac{d^2 r_2}{4\pi}dz_2
\sum
\Psi^*_M(\vec{r}_1,z_1)\Psi_{\gamma}(\vec{r}_1,z_1) \nonumber \\
& \times & \Psi^*_{p^\prime}(\vec{r}_2,z_2)\Psi_{p}(\vec r_2,z_2)
\tilde{J}(\vec b, \vec r_1, \vec r_2, z_1, z_2).
\end{eqnarray}
Here $\vec b$ is the impact parameter of two light-like dipole trajectories
with transverse sizes $\vec r_1$ and $\vec r_2$ respectively and $z_1$, $z_2$
are the longitudinal momentum fractions of the quarks in the dipoles. The
physical picture is that the photon fluctuates into a $q\bar q$ pair, this
is turned into the final meson $M$ by the soft colour interaction $\tilde{J}$,
determined from other reactions \cite{Odderon} and the proton is excited
into an appropriate baryon resonance. The nucleon
and the baryon resonances are treated as quark-diquark dipole systems. The
wave functions automatically take
into account helicity flip at the particle and at the quark level and
produce the correct helicity dependence of $d\sigma/dt$ as $t \to 0$ for
Regge-pole exchange. In elastic hadron-hadron scattering the
increase of the cross sections, together with the shrinking of the
diffractive peak, can be reproduced in this model by suitable scaling of
the hadronic radii. The assumption that the same radial scaling is relevant
for the energy dependence of the Odderon contributions, leads to the
photoproduction cross sections scaling as $(\sqrt{s}/20)^{0.3}$ \cite{Odderon}.

\begin{table}[t]
\centering
\renewcommand{\arraystretch}{1.5}
\begin{tabular}{l c c }
\hline
 $f_2(1270)$  & $\sqrt{s}=7$ TeV & $\sqrt{s}=14$ TeV \\
\hline
    $\frac{d\sigma}{dy}(y=0)$  & 2.4 nb & 2.9 nb \\
    $\sigma_{tot}$ &  29 nb & 37 nb \\
  \hline
\end{tabular}
\caption{Rapidity distribution at $y=0$ and integrated cross sections for the photon-Oderon production channel at the LHC energies.}
\label{tab4}
\end{table}

In what follows we estimate the photon-Odderon contribution (meson photoproduction) to the $f_2(1270)$ exclusive production using the equivalent photon approximation. In this case, the proton-proton cross section can be written as the convolution of the probability of the proton emit a photon with the photon-nucleon cross section producing a resonance ($\gamma p \rightarrow f_2+N$):
\begin{eqnarray}
\sigma_{pp(\gamma p)\rightarrow pRp}(\sqrt{s}) = \int_{Q_{min}^2}^{Q_{max}^2}\int_{x_{min}}^1\frac{d^2n_{\gamma\gamma}}{dQ^2dx}\sigma_{\gamma p \rightarrow R}\,dQ^2dx,\nonumber \\
\,\,
\label{foto}
\end{eqnarray}
where we use the equivalent luminosity spectrum defined in Eq. (\ref{lumi1}) and  using the relation $x=E_{\gamma}/E$. 

In Ref. \cite{Odderon} the authors have computed the photoproduction cross section of $f_2$ meson at the energy $W_{\gamma p}=20$ GeV obtaining the value 21 nb. Moreover, there it has been determined that the energy behavior for Odderon exchange in the photoproduction cross sections scales as $\sigma\,(W_{\gamma p})\propto W_{\gamma p}^{0.3}$. For the photoproduction cross section, we take a simple extrapolation based on the theoretical arguments presented above:
\begin{eqnarray}
\sigma (\gamma p \rightarrow f_2(1270)N) = \sigma(W_0)\left(\frac{W_{\gamma p}^2}{W_0^2} \right)^{0.15},
\end{eqnarray}
where $\sigma(W_0)=21$ nb and $N$ is the nucleon excitation. The energy scale $W_0=20$ GeV is considered, in which the Odderon contribution has been computed \cite{Odderon}.    

 Putting the extrapolation above in Eq. (\ref{foto}), we obtain an estimation of contribution associated to the photon-Odderon production channel. At $\sqrt{s}=20$ GeV the theoretical uncertainty was estimated to be of a factor $2$ \cite{Odderon2}. Similar trend should remain in present case. In the case considered here, the proton is required to break up. This is interesting, since currently all LHC experiments have insufficient  forward coverage, which does not allow a full reconstruction of central exclusive processes.The numerical results for the rapidity distribution at central rapidity and the integrated cross sections are presented in Table IV.  In any case, they are larger than the two-photon channel results (see Table III). Concerning the photoproduction in this channel, it should be noticed that charge asymmetry in the $\pi\pi$ mass spectrum around $f_0$ and $f_2$ mesons may signal the Pomeron-Odderon interference effects as described for instance in \cite{Pire}

As a summary, we have investigated the central diffractive production of mesons $f_0(980)$ and $f_2(1270)$ at the energy of CERN-LHC experiments on proton-proton collisions. For the central diffraction processes we have considered two  non-perturbative Pomeron model to the meson production. In particular, the Donnachie-Landshoff Pomeron model is able to provide the cross section for $J=1,2$ meson states like $f_2(1270)$. The main predictions are the differential cross section for exclusive diffractive $f_0(980)$ meson production, $d\sigma/dy (y=0)\simeq 27$ $\mu$b at the LHC energies as an upper limit and total cross section for exclusive diffractive $f_2(1270)$ meson production, $\sigma\,(f_2)\simeq 1100$ nb.  The theoretical uncertainties are large in such cases, as discussed in text. As a cross check we have computed the cross sections for the exclusive charmonium production and the results are of order magnitude consistent with the Durham and Krakow groups \cite{Durham2010,Krakow2010}.  The double-Pomeron exchange process is the dominant one compared to the two-photon channel which has been considered for sake o completeness. We have also verified the role played by the photon-Odderon production channel.  Namely, we study the implication of high-energy photoproduction of $C=+1$ mesons as $f_2(1270)$ with nucleon excitation through Odderon exchange. We found that such a contribution could be relevant if proton tagging is not imposed.
Concerning the decay channels for the mesons investigated here, the two-pion decay is the dominant one. The branching ratio for $f_2(1270)\rightarrow \pi\pi$ is about 84.8 \%  and for the $f_0(980)$ meson one has $\Gamma (\pi\pi)/[\Gamma (\pi \pi)+\Gamma (K\bar{K})]=0.75$ \cite{PDG}. Actually, this is the verified signal measured at the ALICE experiment that is able to detect the pion pair for double and no-gap events. In particular, in the double gap distribution, the $K_s^0$ and $\rho^0$ are highly suppressed while the $f_0(980)$ and $f_2(1270)$ with quantum numbers $J^{PC}=(0,2)^{++}$ are much enhanced. This enhancement for such states is evidence that the experimental double gap condition used for ALICE selects events dominated by double Pomeron exchange.

\begin{acknowledgments}
The author thanks Wolfgang Ochs and Bernard Pire for useful remarks and comments. The author also acknowledges the hospitality of the Kirchhoff-Institut of the University Heidelberg and the organizers of the {\it WE-Hereaus Summerschool: Diffractive and electromagnetic processes at high energies } (Heidelberg, Sept. 5-9 2011), where this work has started. This research was supported by CNPq, Brazil.
\end{acknowledgments}

\end{document}